\documentclass[11pt]{article}
  \usepackage{amssymb}
   \usepackage{graphicx}
   \usepackage{bm}

\def\ba{\begin{eqnarray}}
\def\bam{\begin{array}}

\def\be{\begin{equation}}

\def\ea{\end{eqnarray}} 
\def\ee{\end{equation}}

\def\fr{\frac}

\def\ha{\frac{1}{2}~}

\def\ts{\textstyle}

\def\1{{\it one}}

\def\2{{\ts{\ha}\!}}

\def\3 {\ts{\frac{1}{3}\!}}
\def\4{\ts{\fr{1}{4}\!}}

\begin{document}

\title{{\bf  On Fast Linear Gravitational Dragging
}}
\author{ D. Lynden-Bell$^{1}$\thanks{email:dlb@ast.cam.ac.uk}, \,
J. Bi\v{c}\'{a}k$^{1,2}$\thanks{email:bicak@mbox.troja.mff.cuni.cz}
\,and\,
J. Katz$^{1,3}$\thanks{email:jkatz@phys.huji.ac.il}
\\
\\ {\it$^1$  Institute of Astronomy, Madingley Road,}
 \\{\it Cambridge CB3 0HA,United Kingdom}
\\
{\it$^2$ Institute of Theoretical Physics, Faculty of Mathematics and Physics,}
\\
{\it Charles University, 180 00 Prague 8, Czech Republic}
\\
{\it$^3$ The Racah Institute of Physics, Edmond Safra Campus,}
\\{\it Givat Ram, 91904 Jerusalem, Israel}}

\maketitle
\abstract
 A new formula is given for the fast linear gravitational dragging of the inertial frame within a 
 rapidly accelerated spherical shell of deep potential. The shell is charged and is electrically
 accelerated by an electric field whose sources are included in the solution.
  
\section*{}

	One hundred years ago in Prague Einstein wrote the paper [1] in which he considered for the first time the dragging of inertial frames. It was within his early variable-velocity-of-light gravitational theory that he calculated the effect of a uniformly accelerated massive shell on a test particle within it. Much effort has since been devoted to his second problem, rotational dragging effects within general relativity, both theoretically and experimentally. Linear gravitational dragging is harder to study since one needs a source for the acceleration which necessarily enters the field equations. This point is emphasised in the model recently analysed by Pfister, Frauendiener and Hengge [2] in which a highly charged and massive spherical shell is linearly accelerated by much smaller oppositely charged clouds. See that paper for further references to linear dragging.
	 In our paper [3] we circumvented the difficulty with the "source of acceleration" by using the special but simple and beautiful solutions of the Einstein-Maxwell equations called conformastats by Synge.  In these, gravitational forces are exactly balanced by electrical forces. The metrics are of the form 
\begin{equation}
ds^2 = V^{-2} dt^2 - V^2 (dx^2 + dy^2 + dz^2), 	 
\end{equation}
where $V = V(x,y,z)\rightarrow1 $ at infinity, and
\begin{equation}
 V^{-3}\nabla^2 V = - \kappa \rho /2 \le 0 ,
\end{equation}
where we use units with $c=1,\kappa = 8\pi G$, and $\nabla^2$ is the flat-space coordinate operator;
this defines the proper density $\rho$, but in terms of the density in coordinate space, $\rho_c$ 
, we have 
\begin{equation}
\nabla^2V=-\kappa \rho V^3/2=-\kappa \rho_c/2.
\end{equation}
The metric (1) is a solution of the Einstein-Maxwell equations with the electric field
\begin{equation}
{\bf E}  = \sqrt{G} \bm{\nabla} V^{-1}=-\sqrt{G} \bm{ \tilde { \nabla} } \ln V,
\end{equation}
where $\bm {\nabla}$  is the coordinate-space gradient and 
$\bm{\tilde  \nabla}$  is the physical    gradient per unit proper length. The charge density $\rho_e$   is equal to the mass density $\rho$    or in standard units  $\rho_e= \sqrt{G} \rho$  (see Appendix C in [3] for more details of these solutions).

In this note we reconsider the problem of linear dragging analysed in [3] by giving a new and physically more appropriate interpretation of our solution.
	Our problem concerns the acceleration of an uncharged test particle inside a small spherical shell of radius $b$, mass $m$, and charge $q= \sqrt{G} m$,  which sits with electrical and gravitational forces in balance in the field of a large mass $M$ of charge $Q=\sqrt{G}M$, see figure 1.  

\begin{figure}[htbp]
\begin{center}
\includegraphics[width=8cm]{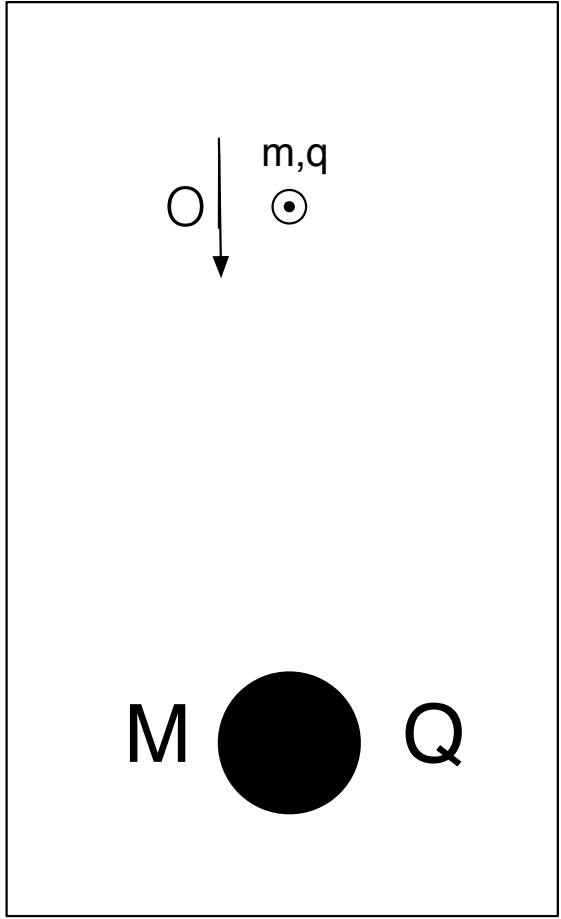}
\caption{The observer O falls towards $M$.
	     He sees the uncharged test particle
	     within the charged mass shell to fall
	            less rapidly than he does.}
\label{fig.1}
\end{center}
\end{figure}
An uncharged freely-falling observer sees the charged shell to be rapidly accelerated upward by the electrical field of the large mass; but the test particle within the shell does not fall as rapidly as the observer, who attributes the difference to the dragging of the local inertial frame by the small shell which has a deep potential well, as $m/b$ is not small.
	In our paper [3] the small shell had a uniform density in $ x,y,z$ space and in units with
$G = 1$,  $V$  is given  by $V= 1+M/R+m/r$  for $r\ge b$, where $r$ is coordinate distance from the origin inside the shell, and $R$ is coordinate distance from the mass $M$ at $(0,0,-Z)$. What we did not consider was that  $R$ varies across the small shell from $Z-b$ to $Z+b$ so a uniform coordinate density leads to a non-uniform proper density that varies by a factor $[(1+MbZ^{-2})/ (1-MbZ^{-2})]^3$,  thus giving us a loaded shell in proper density. In estimating the dragging, the field due to this loading-of-the-dice should be allowed for. We may also question whether the coordinate sphere is  actually spherical, since departures from sphericity might also give rise to internal fields. 
	 The metric in which our shell sits is  given by (1) in which $V=1+M/R+m/b$ for $r<b$. Evidently the spatial metric is a conformal transformation of flat space. A simple way of eliminating the loading problem is to take our shell to be a conformally transformed spherical shell of uniform proper density rather than a coordinate-spherical shell of uniform coordinate density. True infinitesimal distances are given by $Vdx,Vdy,Vdz$. Near the axis $V$ is primarily dependent on $z$, so inside the small shell we rescale the proper length space by the constant factor $V_0$ and write $dx'=Vdx/V_0$, where $V_0=1+M/Z+m/b$. Then in the region of the shell the transformation from the $x,y,z$ coordinates to the uniformly scaled true distances is given by $z'=\int_0^zV/V_0dz $ etc., so
\begin{equation}
	x'=x(1-z/\lambda) ;\quad  y'=y(1-z/\lambda) ;\quad  z'=z[1-z/(2\lambda)],
\end{equation}
where $ \lambda=Z^2V_0/M>Z$.	
Since $b$ is small compared to $Z$, all these corrections are small inside the shell. 
We now imagine a true sphere in the rescaled distances given by 
\begin{equation}
x'^2+y'^2+z'^2=b^2,
\end{equation}
and we put a uniform proper density in the shell of radius $b$ and small thickness.
When we apply our conformal transformation the sphere will become 
\begin{equation}
x^2(1-2z/\lambda)+y^2(1-2z/\lambda)+z^2(1-z/\lambda)=b^2,
\end{equation}
so in our coordinate space it will not be quite spherical but have a radius $r+\delta r$,where
\begin{equation}
\delta r=(b/\lambda) \cos{\theta}[1-(1/2)\cos^2{\theta}]= (b/\lambda)[(7/10)P_1-(1/5)P_3],
\end{equation}
where the $P_l$ are Legendre polynomials and $z=r \cos{\theta}$.
	Not only is the shell no longer quite spherical but also its density will be changed by the cube of the conformal factor  as in equation (3). As the thickness also changes by the inverse of the conformal factor the surface density of mass in coordinate space $\sigma_c$ will no longer be  $m/(4\pi b^2)$ but will vary as $[1+mb^{-1}+MZ^{-1}(1-z/Z)]^2/V_0^2$, i.e. as $1-2b\lambda^{-1}\cos{\theta}$.
Thus our proper sphere of uniform proper density is neither quite spherical nor uniform in coordinate space. The potential $V_s$ of such a quasi-sphere can be calculated as in Newtonian theory because we have to solve the flat-space equation (3). Hence inside we have
\begin{equation}
V_s=\sum{ A_l(r/b)^lP_l},
\end{equation}
and outside
\begin{equation}
V_s=\sum B_l(b/r)^{l+1}P_l.
\end{equation}
These must be equal on the surface $r=b+\delta r$ and their radial derivatives there must be related by
\begin{equation}
-(\partial V_s/\partial r)_{out}+(\partial V_s/\partial r)_{in}=4\pi\sigma_c=mb^{-2}(1-2b\lambda^{-1} \cos{\theta}). 
\end{equation}
The only complication is that $\delta r$ depends on $\theta$ which generates terms up to $P_3$ in the potential. Equating the potentials inside and outside on the surface we find
\begin{equation}
m/b+A_1P_1+A_3P_3=m/b+(m/\lambda)((7/10)P_1-(1/5)P_3)+B_1P_1+B_3P_3,
\end{equation}
and from equation (11)
\begin{equation}
(m/b)(1-2b/\lambda P_1)=(m/b)-(2m/\lambda)((7/10)P_1-(1/5)P_3)+(A_1+2B_1)P_1+(3A_3+4B_3)P_3.
\end{equation}
For $l>0$ the $A_l$ and the $B_l$ are $O(m/Z)$ and we discard terms of order $(m/Z)^2$. Using the independence of the $P_l$ we equate their coefficients and solve for the $A_l$ and the $B_l$ obtaining,
 \begin{equation}
A_1=-(2/3)m/\lambda;A_3=(2/105)m/\lambda;B_1=(1/30)m/\lambda;B_3=-(19/105)m/\lambda.
\end{equation}
The potential of our shell of mass $m$ with uniform proper density, together with the large mass  $M$, is (within the shell)
 \begin{equation}
      V=1+M/R+m/b+(m/\lambda)[-(2/3)(r/b)P_1+(2/105)(r/b)^3P_3],    
\end{equation}
so at the centre $\partial V/\partial z=-M/Z^2-(2/3)m/(\lambda b)=-MZ^{-2}V_0^{-1}[1+M/Z+(5/3)m/b]$.
It is the final square bracketed term in equation (15)  that amends our former result.

Three possible measures of acceleration considered earlier are:
1) coordinate acceleration $g_c=d^2z/dt^2=V^{-5}\partial V/\partial z$,
2)  proper length acceleration in proper time $g_p=d/d\tau(Vdz/d\tau)=V^{-2}\partial V/\partial z$,      and 3) proper length acceleration in the universal $t$-time  $g_u=d/dt(Vdz/dt)=V^{-4}\partial V/\partial z$;  here we used the equations of geodesic motion to get the $V^{-n} \partial V/\partial z$ terms and $\tau$ is the proper time measured on the accelerating body. Each $g$ depends on $V$ itself, not just its derivative, so for the test particle within the shell this depends on its $z$. For simplicity we shall take it to start at the shell's centre $z=0$.  Since the observer is too far away to be influenced by the small sphere, for him $V$ is just $(1+M/R)$ with $R= Z$; 
so for the observer $g_{co}=-MZ^{-2}/(1+M/Z)^5$ and $g_{po}$ and $g_{uo}$ are given by replacing the power 5 by 2 or 4 respectively. By contrast for the test particle within the shell,
\begin{equation}
g_{cp}=-MZ^{-2}[1+(5/3)\mu)]/V_0^{n+1},
\end{equation}
where $\mu=m/[b(1+M/Z)]$,
with $n=5$, and $n=2,4$ for the other accelerations. The ratios of the accelerations of the test particle to those of the observer are all of the form, 
\begin{equation}
g_{p}/g_{o}=[1+(5/3)\mu]/(1+\mu)^{n+1}.
\end{equation}
There is no value of $\mu$ for which this ratio is one, so on all measures dragging is always there.
Were we to consider a race towards the large mass with the observer and the test particle being the competitors, both starting at $z=0$ and aiming for  $z= -Z$, then we should compare their starting accelerations via $g_c =d^2z/dt^2$    rather than the other measures. Using that measure of who is getting off to the best start but converting it to his own time, our observer sees a dragging of the test particle by the accelerating small shell of
\begin{equation}
D=(1+M/Z)^2[g_{cp}-g_{co}]=MZ^{-2}(1+M/Z)^{-3}[(1+\mu)^6-1-(5/3)\mu)]/(1+\mu)^6).
\end{equation}
We may re-express this result in terms of the excess depth gravitational potential due to the sphere,
$\delta=\ln(1+\mu)$, and $\alpha$ the sphere's acceleration as seen by  the observer.
\begin{equation}
D=\alpha[1-e^{-6\delta}((5/3)e^{\delta}-2/3)]\rightarrow (13/3)\alpha \delta,
\end{equation}
where the last expression is for small $\delta$.
This is our final result for the dragging observed.

\end{document}